\documentclass[final,5p,times,twocolumn, sort&compress]{elsarticle}

\usepackage{amssymb,color}
\usepackage{graphicx}
\usepackage{epstopdf}

\graphicspath{{figs/}}

\journal{Journal of the Mechanics and Physics of Solids}

\begin{document}

\begin{frontmatter}



\title{The Effects of Free Edge Interaction-Induced Knotting on the Buckling of Monolayer Graphene}


\author[shu]{Hao-Yu Zhang}

\author[shu]{Jin-Wu Jiang\corref{cor}}
\ead{jiangjinwu@shu.edu.cn}

\author[shu]{Tienchong Chang}

\author[shu]{Xingming Guo}

\author[bu]{Harold S. Park}

\cortext[cor]{Corresponding author}

\address[shu]{Shanghai Institute of Applied Mathematics and Mechanics, Shanghai Key Laboratory of Mechanics in Energy Engineering, Shanghai University, Shanghai 200072, People's Republic of China}

\address[bu]{Department of Mechanical Engineering, Boston University, Boston, MA 02215, USA}

\begin{abstract}

Edge effects play an important role for many properties of graphene.  While most works have focused on the effects from isolated free edges, we present a novel knotting phenomenon induced by the interactions between a pair of free edges in graphene, and investigate its effect on the buckling of monolayer graphene.  Upon compression, the buckling of graphene starts gradually in the form of two buckling waves from the warped edges. The collision of these two buckling waves results in the creation of a knot structure in graphene. The knot structure enables the buckled graphene to exhibit two unique post-buckling characteristics.  First, it induces a five-fold increase in graphene's mechanical stiffness during the buckling process.  Second, the knotted structure enables graphene to exhibit a mechanically stable post-buckling regime over a large (3\%) compressive strain regime, which is significantly larger than the critical buckling strain of about 0.5\%.  The combination of these two effects enables graphene to exhibit an unexpected post-buckling stability that has previously not been reported. We predict that numerical simulations or experiments should observe two distinct stress strain relations for the buckling of identical graphene samples, due to the characteristic randomness in the formation process of the knot structure.

\end{abstract}


\end{frontmatter}



\section{Introduction}

Graphene is a quasi two-dimensional (2D) honeycomb lattice structure that exhibits extremely high in-plane stiffness~\cite{LeeC2008sci} but very small bending stiffness~\cite{OuyangZC1997,TuZC2002,ArroyoM2004,LuQ2009}. The quasi 2D nature of graphene is the origin for many of the interesting phenomena involving graphene, including edge effects and buckling instability, which are of relevance to the present work. 

For the buckling instability, Euler buckling theory~\cite{TimoshenkoS1987} states that the critical compression strain, above which graphene is buckled, is inversely proportional to the in-plane stiffness $C_{11}$ and is proportional to the bending stiffness $D$; i.e., $\epsilon_c \propto D/C_{11}$. According to Euler buckling theory, the critical strain for graphene is very small. Consequently, the buckling process can be induced by very weak external disturbances such as thermal expansion~\cite{BaoW2009nn}. As a result of the buckling phenomenon, graphene is bent or folded with a finite curvature, which can be used to manipulate many physical properties in graphene~\cite{CongC2014nc}. As a result, the buckling of graphene has attracted intensive research interest in past few years~\cite{LuQ2009ijam,PatrickWJ2010jctn,SakhaeePA2009cms,PradhanSC2009cms,PradhanSC2009plsa,FrankO2010acsnn,FarajpourA2011pe,TozziniV2011jpcc,RouhiS2012pe,GiannopoulosGI2012cms,Neek-AmalM2012apl,ShenH2013apl}. Besides graphene, a group of other quasi 2D materials, eg. MoS$_2$ or black phosphorus, also have small critical buckling strains because the bending stiffnesses for these atomically thin materials are also very small~\cite{JiangJW2014mos2buckling,JiangJW2015reviewgramos2}.

As another result of graphene's 2D nature, edge effects play an important role on its physical properties. Based on the Brenner atomic potential~\cite{brennerJPCM2002} and the finite element method, it was demonstrated that graphene's free edges can become warped due to the compressive edge stress~\cite{ShenoyVB}. The warping amplitude decays exponentially from the edge into the center; i.e., the height (z) of the warped configuration is $z\propto e^{y/l_c}$ with $l_c$ as the critical penetration depth. The critical penetration depth can be viewed as the size of the warped edge region.  For narrow graphene nanoribbons, the size of the edge region can be comparable to or larger than the central region.

If the size of the edge region in graphene nanoribbons is sufficiently large, the free edges dominate most of graphene's physical properties.  Edge reconstructions have been observed experimentally~\cite{MhairiHG2008nn}, which can be attributed to the thermal energy localized by the edge vibrations~\cite{JiaX2009sci,EngelundM2010prl}. Edge vibrations were also found to be responsible for the larger energy dissipation in graphene nanomechanical resonators~\cite{KimSY2009nl,JiangJW2012jap}. It was found that edge effects are the dominant factor for the friction between neighboring nanotubes in multi-wall carbon nanotubes~\cite{GuoZ2011prl}, and a piece of graphene can be driven from a softer regime to the stiffer regime due to the edge effect~\cite{ChangT2015prl}. While we have listed just a few examples here, free edges also have a strong effect on other physical properties in graphene (for review, see eg. Ref.~\cite{CastroNAH}).

Although edge effects on the mechanical properties in graphene have been extensively studied, the edge effect on buckling has not been examined to-date. Furthermore, free edges almost always are present in pairs. However, in the aforementioned works, each free edge makes an independent contribution to those mechanical properties in graphene. If the width of the graphene is comparable with twice the critical penetration depth $l_c$, there should be a strong correlation and interactions between the pair of free edges. The effect from a pair of correlated edges on the mechanical properties of graphene has not been studied yet. We thus investigate the effect from a pair of correlated edges on the buckling phenomenon in graphene.

In this paper, we investigate the buckling process for graphene with a pair of free edges. Different from the usual abrupt buckling mode, we find that graphene is gradually buckled starting from the free edges if the two edges are warped in opposite directions. The gradual buckling is due to the formation of a knot structure that results from the collision of the buckling waves from the two edges. There are four major features brought by the knotting effect. (1) Graphene with knotted structure has a much higher mechanical stiffness than graphene without knotting during the buckling process. (2) It is more difficult to buckle narrower graphene nanoribbons with the knotted structure as the knot is stronger in narrower graphene. (3) As a result of the randomness in the knotting phenomenon, we predict that numerical simulations or experiments should observe two different buckling processes even for identical graphene samples with free edges. (4) The knot is formed by the collision of  buckling waves from the two free edges, and the knot structure will be unknotted if the compressive strain is larger than a critical unknotting strain value. After unknotting, all graphene with different boundary conditions have the same final buckled structure.

\section{Simulation details}

\begin{figure}
  \begin{center}
    \scalebox{1.0}[1.0]{\includegraphics[width=8cm]{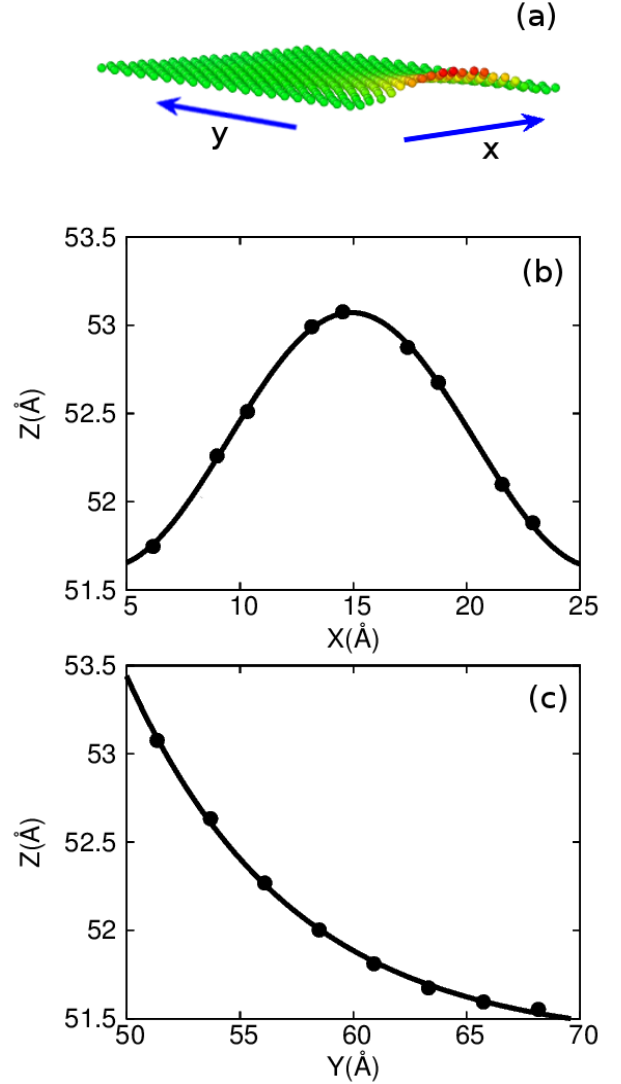}}
  \end{center}
  \caption{(Color online) Warped configuration at 1~K of a free edge in graphene of dimension $30\times 80$~{\AA}. Half of the system is shown in the figure, while the other half (with another warped edge) is not shown. (a) Perspective view of the warped edge. The warped shape is described by the function $z(x,y)=z_0 + A\sin(\pi x/L)e^{-y/l_c}$. (b) z-position for atoms at $y=y_{\rm min}$. (c) z-position for atoms at the middle plane $x=15$~{\AA}. The color is with respective to the z-position of each atom. Graphene is compressed or stretched in the x-direction.}
  \label{fig_cfg_one_edge}
\end{figure}
  
The interaction between carbon atoms in graphene is described by the second generation Brenner potential~\cite{brennerJPCM2002}. For stretching or compression, the edges of the graphene in the strain direction, i.e. the +x and -x edges in Fig.~\ref{fig_cfg_one_edge}~(a), have prescribed motion in the strain direction only, while free boundary conditions (FBC) are used in the out-of-plane direction. Before tension or compression, the system is thermalized to a targeted pressure and temperature within the NPT (i.e. the particles number N, the pressure P and the temperature T of the system are constant) ensemble for 200~ps. The Nos\'e-Hoover~\cite{Nose,Hoover} thermostat is used for maintaining constant temperature and pressure. After thermalization, graphene is stretched or compressed in the x-direction in Fig.~\ref{fig_cfg_one_edge}~(a) by uniformly deforming the simulation box in this direction, while the structure is allowed to be fully relaxed in lateral directions during mechanical loading. The standard Newton equations of motion are integrated in time using the velocity Verlet algorithm with a time step of 1~{fs}. Molecular dynamics (MD) simulations are performed using the publicly available simulation code LAMMPS~\cite{PlimptonSJ,Lammps}. The OVITO package was used for visualization~\cite{ovito}.

\section{An isolated edge}

\subsection{Warped Configuration}

It has been demonstrated that free edges are warped due to the compressive edge stress in graphene~\cite{ShenoyVB}. A typical warped edge configuration is illustrated in Fig.~\ref{fig_cfg_one_edge}~(a). The dimension of the graphene is $30\times 80$~{\AA}. The two ends in the x-direction are fixed, while FBC is applied in the y-direction. Only half of the system is shown, while the other warped edge is not displayed. The structure is relaxed at 1.0~K. The warping amplitude decays exponentially from the free edge into the center. Fig.~\ref{fig_cfg_one_edge}~(b) and (c) show that the height (z) of each atom can be well described by the function $z(x,y)=z_0 + A\sin(\pi x/L)e^{-y/l_c}$, where $L=30$~{\AA} is the length of graphene along the x-direction. Fitting parameter $A$ is the warping amplitude, and $l_c=7.3$~{\AA} is the critical penetration depth of the warping edge.

\begin{figure}
  \begin{center}
    \scalebox{1.0}[1.0]{\includegraphics[width=8cm]{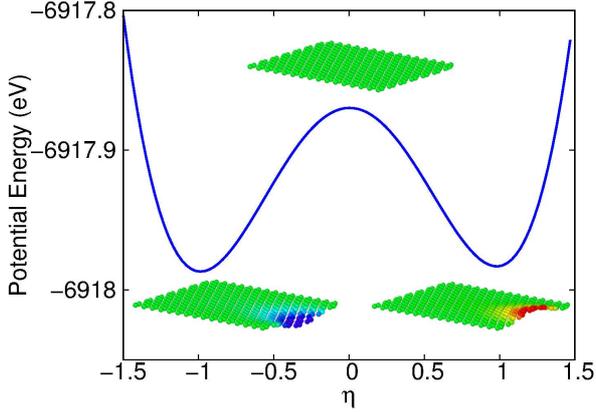}}
  \end{center}
  \caption{(Color online) Potential energy for graphene nanoribbon described by $\textbf{R}=\frac{1-\eta}{2}\textbf{R}_{-} + \frac{1+\eta}{2}\textbf{R}_{+}$, in which $\eta$ is an evolving parameter. Two lower insets correspond to $\eta=\pm 1$, while the top inset is the structure for $\eta=0$.}
  \label{fig_potential_strain_0.0}
\end{figure}

An isolated free edge can be warped either in the +z or -z direction, whose structures are denoted by $\eta=\pm 1$ in Fig.~\ref{fig_potential_strain_0.0}, and whose corresponding configurations are displayed as the two lower insets in the figure. Fig.~\ref{fig_potential_strain_0.0} shows that these two warping configurations have the same potential energy, as they are symmetric with respective to the z=0 plane. It means that the probability for an isolated free edge to warp in the +z direction is the same as -z direction.

\subsection{Thermally Induced Flipping of the Warped Edge}

While the results in Fig.~\ref{fig_potential_strain_0.0} were for a single temperature, it is intuitive that as temperature increases, the thermal vibration energy may become large enough to flip the warping direction of the free edge; i.e., the +z-warping edge can be flipped into the -z-warping edge, and vice versa. To determine the critical temperature, we plot in Fig.~\ref{fig_potential_strain_0.0} the potential energy curve for the graphene structure evolved by parameter $\eta$. The fact that we are computing an energy landscape implies that these, and subsequent potential energy surface calculations are performed at 0~K. The graphene configuration with $\eta=-1$ corresponds to the structure shown in the left bottom inset (denoted by $\textbf{R}_{-}$), where the edge is warped downward. Only half of the structure is shown, while the other half (not shown in the inset) remains unchanged during the $\eta$ evolution. The configuration with $\eta=+1$ corresponds to the structure shown in the right bottom inset (denoted by $\textbf{R}_{+}$), where the free edge is warped upward. A general graphene configuration is determined by parameter $\eta$ following the formula, $\textbf{R}=\frac{1-\eta}{2}\textbf{R}_{-} + \frac{1+\eta}{2}\textbf{R}_{+}$. The top inset displays the graphene configuration corresponding to $\eta=0$.

From the potential energy curve, the two configurations with $\eta=\pm 1$ are two stable states with the same potential; i.e., this is a bistable system. The atomic color is with respective to the z-coordinate of each atom. The potential barrier between configurations $\eta=\pm 1$ is $\Delta V=V_{\eta=0} - V_{\eta=-1}=0.117$~eV. The number of atoms in the warped edge regime is $N_E=4\times (W\times l_c/s_0) = 4\times (24.0\times7.3/10.48) = 64$, where $s_0=10.48$~\AA$^2$ is the area for one cell containing four carbon atoms. The potential energy barrier per atom is thus about $\Delta V / N_E =1.83$~{meV/atom}. The probability to overcome this energy barrier at finite temperature T is proportional to $e^{-\Delta V/k_BT}$, so the critical temperature can be extracted as $T_C=\Delta V/k_B=18.3$~K. This critical temperature means that, for $T>T_C$, the free edge can be driven from configuration with $\eta=-1$ to the configuration with $\eta=1$ purely by the thermal vibrations, so these two configurations can switch between each other by thermal vibrations.

\begin{figure}
  \begin{center}
    \scalebox{1.0}[1.0]{\includegraphics[width=8cm]{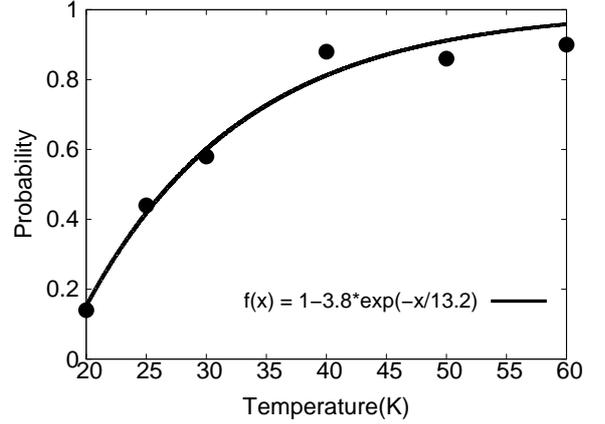}}
  \end{center}
  \caption{(Color online) Temperature dependence for the flipping probability of one isolated warped edge in a graphene nanoribbon of dimension $30\times 80$~{\AA}.}
  \label{fig_unstable_tem}
\end{figure}

To verify the above potential barrier argument, we perform MD simulations for the warped free edge in a graphene nanoribbon of dimension $30\times 80$~{\AA}. We ran 50 simulations for this graphene sample at each temperature.  Each simulation is performed using a different random velocity distribution, while all other simulation conditions remain unchanged. The warping direction of the free edge is flipped in many of the simulations, based upon which the flipping probability is calculated. Fig.~\ref{fig_unstable_tem} shows the temperature dependence for the flipping probability of one isolated warped edge. The warping direction of the free edge can be flipped by thermal vibrations for temperatures above 20~K. It means that the thermal vibrations for $T>20.0$~K are able to overcome the potential energy barrier of the warped free edge in Fig.~\ref{fig_potential_strain_0.0}, resulting in the flipping of the warped free edge.  In contrast, there is almost no flipping of the warped free edge for temperatures below 20~K, which is very close to the critical temperature of 18.3~K for the warped edge in Fig.~\ref{fig_potential_strain_0.0}.

\section{A Pair of Edges and the Knotting Effect}

\subsection{Structure for Interacting Edge Pair}

\begin{figure}
  \begin{center}
    \scalebox{1}[1]{\includegraphics[width=8cm]{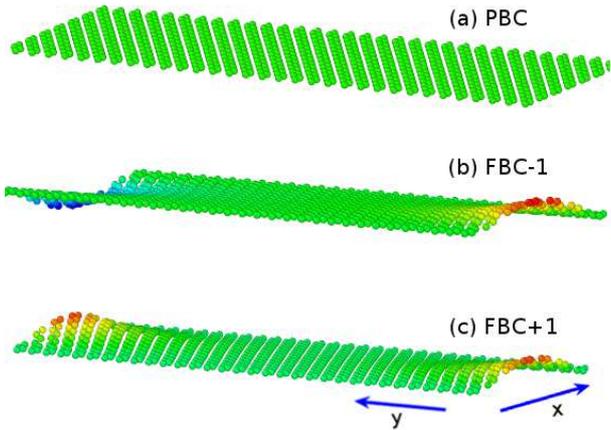}}
  \end{center}
  \caption{(Color online) Structure of a pair of free edges at 1~K. (a) PBC case. Graphene is flat, with PBC in the y-direction. (b) FBC-1 case. The two edges are warped in the opposite direction. (c) FBC+1 case. The two edges are warped in the same direction.}
  \label{fig_sin_e_structure}
\end{figure}

We have discussed above the structure of an isolated free edge, but free edges normally show up in pairs, which we now consider.  Fig.~\ref{fig_sin_e_structure} shows three different configurations for a graphene nanoribbon of dimensions $30 \times 80$~{\AA} where the two shorter edges are free, and where the two longer edges are fixed.  Fig.~\ref{fig_sin_e_structure}~(a) shows that graphene at 1~K has a flat configuration if periodic boundary conditions (PBC) are applied in the y-direction. Fig.~\ref{fig_sin_e_structure}~(b) and (c) illustrate two possible edge structures for FBC along the y-direction. The warping directions of the pair of edges are in opposite directions in Fig.~\ref{fig_sin_e_structure}~(b), which will be referred to as the FBC-1 configuration; while the warping of the pair of free edges is in the same direction in Fig.~\ref{fig_sin_e_structure}~(c), which is referred to as the FBC+1 configuration.

\subsection{Knotting Effect on Buckling}

\subsubsection{Identification of Knotting Effect from Stress-Strain Relationship}

A thin plate (like graphene) will buckle under a sufficiently large compressive loading~\cite{TimoshenkoS1987}. The buckling phenomenon is typically described in two stages. First, external work is done to compress the plate, and the energy is accumulated as compressive strain energy in the plate. The planar structure for graphene is kept in this process. Second, after the compressive strain reaches a critical value $\epsilon_c$, and graphene's planar structure becomes unstable, buckling happens abruptly, where the compressive energy inside the planar structure is fully converted into the bending energy of the buckled structure. The value of the critical buckling strain can be determined by equating the compressive strain energy of the plate just prior to buckling and the bending energy in the buckled structure.

We note one important condition in the Euler buckling theory is that the plate is in a planar configuration at the beginning of the mechanical compression. As a result, there is no bending energy in graphene during the pre-buckling stage.  However, for graphene with FBC, the free edge is warped into the non-planar shape $z(x,y)=z_0 + A\sin(\pi x/L)e^{-y/l_c}$, so the bending energy coexists with the compressive energy in graphene even in the pre-buckling stage with $\epsilon<\epsilon_c$. As a result, the buckling process may be quite different due to the warped free edges in graphene.

\begin{figure}
  \begin{center}
    \scalebox{1.0}[1.0]{\includegraphics[width=8cm]{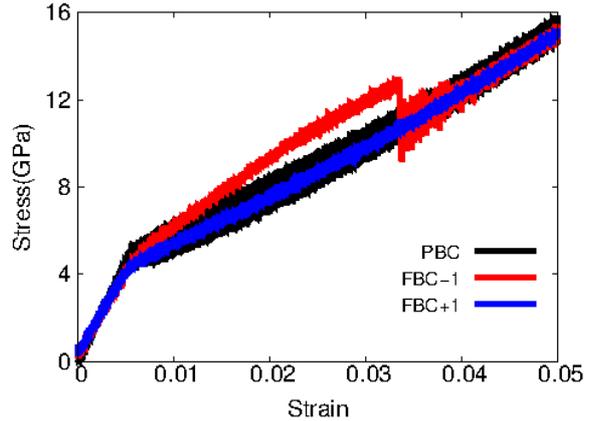}}
  \end{center}
  \caption{(Color online) Stress-strain curves for the compression of a graphene nanoribbon of dimension $30\times 80$~{\AA} at 1.0~K.}
  \label{fig_width80}
\end{figure}

We thus simulate the compressive response of graphene with length $L=30$~{\AA} in the x-direction and width $W=80$~{\AA} in the y-direction.  Fig.~\ref{fig_width80} compares the stress-strain curves at 1.0~K for graphene with PBC, FBC-1, and FBC+1 configurations. For graphene with PBC, the stress-strain curve is as expected; i.e., the curve changes its slope at the critical buckling strain $\epsilon_c=0.0052$, at which point the structure is buckled abruptly. 

There are two different stress-strain curves (red and blue lines) in Fig.~\ref{fig_width80} corresponding to the buckling of graphene with FBC. Graphene with FBC+1 configuration has a similar stress-strain relation as the PBC configuration. However, there are several distinct features in the stress-strain relation of the FBC-1 case. First, the slope of the stress-strain curve changes gradually before the critical unknotting strain $\epsilon_{u}=0.0336$, indicating a gradual buckling mode of the graphene. Different from the standard critical buckling strain $\epsilon_c$, $\epsilon_u$ is a new critical strain, above which the knot structure is unknotted as shown in the following. Second, the achievable stresses are larger for the FBC-1 case, which indicates that graphene with FBC-1 configuration has a much higher mechanical stiffness during the buckling process. Third, for strain $\epsilon>\epsilon_u$, the stress-strain curve of FBC-1 case jumps down and coincides with the PBC and FBC+1 cases. The distinct stress-strain relation indicates some novel effects in the buckling of graphene with FBC-1 configuration, two of which we highlight now.

\begin{figure}
  \begin{center}
    \scalebox{1.0}[1.0]{\includegraphics[width=8cm]{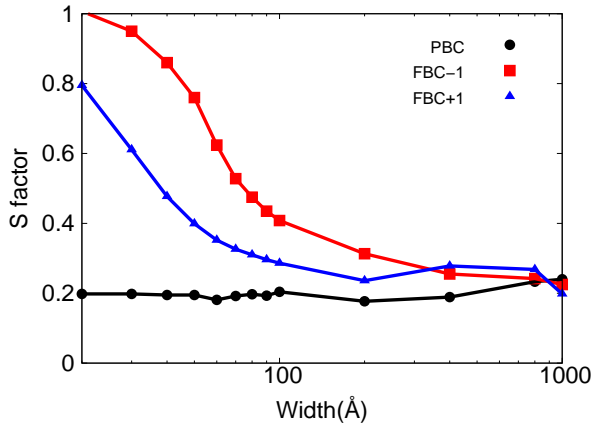}}
  \end{center}
  \caption{(Color online) Width dependence for the S factor of graphene buckling at 1.0~K. The length of the graphene is 30~{\AA}.}
  \label{fig_S}
\end{figure}

First, to provide a quantitative description for the buckling process, we compute the S factor based on the stress-strain curve. The S factor is useful in capturing the buckling effect on the stiffness of the material~\cite{CoulaisC2015prl}, and is defined as $S=\frac{Y_f}{Y_i}$, with $Y_i$ and $Y_f$ as the Young's modulus before and after buckling at the critical strain $\epsilon_{c}$, respectively. The S factor is usually smaller than 1, because the stiffness is reduced by buckling. Fig.~\ref{fig_S} shows the S factor for the buckling of graphene with width $W\in[20, 1000]$~{\AA}. The S factor for graphene with PBC configuration is the lowest one, about 0.2, so the stiffness is greatly reduced by the buckling of graphene with PBC. The S factor is also width independent for graphene with PBC. The S factor in graphene with FBC-1 case is the largest one among all of the three configurations.  In particular, the S factor for FBC-1 is close to 1 for narrow graphene with widths $W<50$~{\AA}, which suggests that the stiffness of the graphene with FBC-1 is essentially unaffected by buckling.  In other words, the mechanical stiffness for graphene with FBC-1 configuration is nearly five times larger than the stiffness of graphene with PBC.  For wide graphene with width $W>80$~{\AA}, the edge effect becomes negligible and the S factors for graphene with PBC, FBC-1, and FBC+1 configurations are quite similar.

Second, as can be seen in Figs.~\ref{fig_width80} and~\ref{fig_S}, the knotting effect enables graphene to show a fairly stable, post-buckling regime whose duration of about 3\% compressive strain as seen in Fig.~\ref{fig_width80} is nearly 6 times larger than the elastic strain that graphene undergoes before buckling.  Therefore, not only can graphene sustain significantly more compressive strain after buckling due to the knotting, it is also very mechanically stable, particularly if the width is smaller than about 80~\AA, as shown in Fig.~\ref{fig_S}.  Together, these effects demonstrate a new post-buckling stability in graphene that has not previously been reported.

\subsubsection{Illustrating the Knotting Effect During Buckling}

\begin{figure}
  \begin{center}
    \scalebox{1}[1]{\includegraphics[width=8cm]{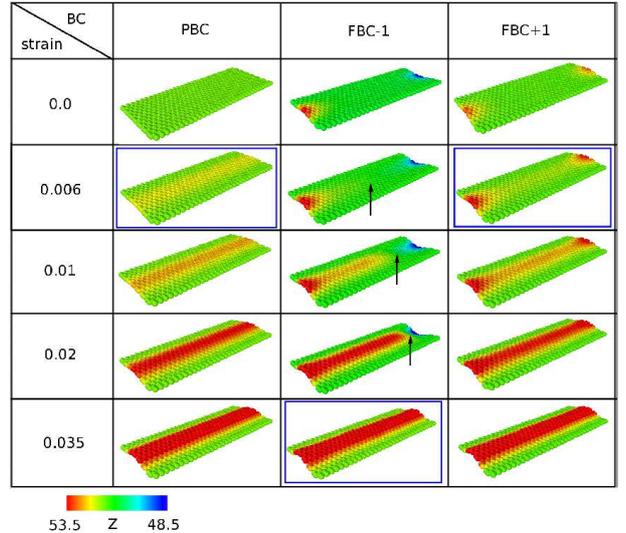}}
  \end{center}
  \caption{(Color online) MD snapshot for the buckling processes at 1.0~K of graphene with dimension $30\times 80$~{\AA}. Left: graphene with PBC. Right: graphene with FBC+1. Middle: graphene with FBC-1. The knot in graphene with FBC-1 configuration is depicted by the black arrow. }
  \label{fig_buckling_all}
\end{figure}

According to the above discussions based on the stress-strain relations, free edges can enhance graphene's ability to resist buckling, particularly in graphene with the FBC-1 configuration. To explicitly disclose the differences in the buckling process, we show in Fig.~\ref{fig_buckling_all} some typical MD snapshots for the buckling process of graphene at 1~K with PBC, FBC-1, and FBC+1 configurations. For graphene with PBC (left), the structure is buckled abruptly at strain $\epsilon_c=0.0052$. For graphene with FBC+1 (right), the buckling starts from the two free warped edges. The edge buckling waves propagate into the interior region. Graphene is buckled after these two buckling waves meet in the central region at almost the same critical strain as PBC case (i.e., $\epsilon_c=0.0052$). The buckled structure for FBC+1 case after $\epsilon>0.02$ is the same as PBC case in the left panel, which explains why graphene with PBC and FBC+1 configurations have similar stress-strain curves just after the critical buckling strain in Fig.~\ref{fig_width80}. 

For graphene with the FBC-1 configuration (middle), the structure also buckles gradually, starting with the propagation of waves propagating in from the free warped edges.  However, different from the FBC+1 case, a stable knot structure is formed in the center of the graphene sheet after the collision of these two edge buckling waves at strain of 0.006. Upon application of additional force, the knot propagates towards one of the free ends.  This knotting configuration enhances the structure's mechanical stiffness during buckling; i.e., higher stress is observed for FBC-1 in Fig.~\ref{fig_width80}. The knotting structrue is unknotted at the critical unknotting strain $\epsilon_u=0.0336$, leading to the final buckled structure. This final buckled structure is the same as the buckled structure for graphene with PBC and FBC+1 configurations. Hence, all of these three stress-strain relations in Fig.~\ref{fig_width80} fall onto one curve after $\epsilon>\epsilon_u$.

\subsubsection{Potential Energy Analysis for Knotting Effect}

\begin{figure}
  \begin{center}
    \scalebox{1.0}[1.0]{\includegraphics[width=8cm]{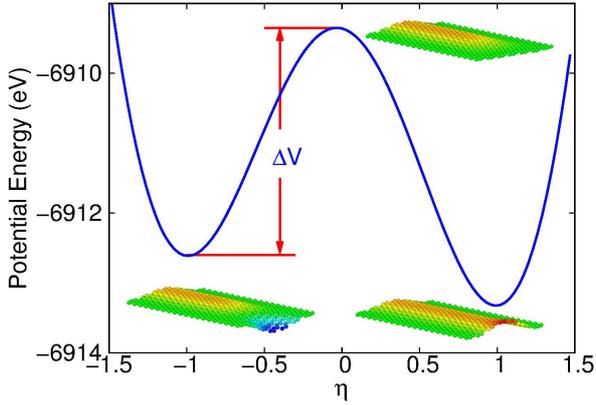}}
  \end{center}
  \caption{(Color online) The potential energy curve of a knotting configuration at strain $\epsilon=0.02$ for graphene of dimension $30\times 80$~{\AA}. Graphene with FBC-1 configuration is compressed and a a knot is formed at strain $\epsilon=0.02$. The configuration is evolved by parameter $\eta$ via $\textbf{R}=\frac{1-\eta}{2}\textbf{R}_{-} + \frac{1+\eta}{2}\textbf{R}_{+}$, where $\textbf{R}_{\pm}$ corresponds to the two configurations in the lower insets, denoted by $\eta=\pm 1$.}
  \label{fig_potential_strain_0.02}
\end{figure}

We now provide a potential energy analysis for the knotting effect on the graphene buckling. Fig.~\ref{fig_potential_strain_0.02} shows the potential energy curve for a knotting configuration at strain $\epsilon=0.02$; i.e., the graphene with FBC-1 configuration is compressed and a knot is formed at strain $\epsilon=0.02$. The x-axis $\eta$ evolves the structure via $\textbf{R}=\frac{1-\eta}{2}\textbf{R}_{-} + \frac{1+\eta}{2}\textbf{R}_{+}$. The structure with $\eta=-1$ corresponds to the structure shown in the left bottom inset ($\textbf{R}_{-}$), which is the knotting structure. Only half of the structure is displayed here, as the other half is not changed during the evolving process. The graphene configuration with $\eta=+1$ corresponds to the structure shown in the right bottom inset ($\textbf{R}_{+}$), which is a more stable structure with lower potential energy. This is the structure after the knot is unknotted.  The top inset illustrates the configuration with $\eta=0$. After the knot is unknotted, the structure transforms from $\textbf{R}_{\eta=-1}$ to $\textbf{R}_{\eta=0}$. For unknotting to occur, external work needs to be done to overcome the potential energy barrier $\Delta V=V_{\eta=0}-V_{\eta=-1}$. 

\begin{figure}
  \begin{center}
    \scalebox{1.0}[1.0]{\includegraphics[width=8cm]{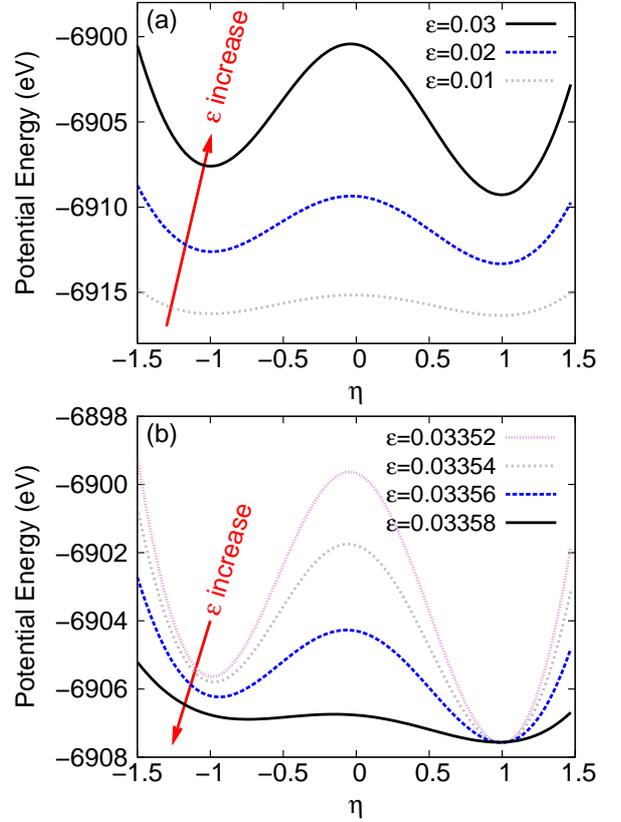}}
  \end{center}
  \caption{(Color online) Potential energy curve of the knotting at different compressive strains $\epsilon$ for graphene of dimension $30\times 80$~{\AA}. (a) Strain is smaller than 0.033. The potential energy curve becomes higher for larger strain. (b) Strain is larger than 0.033. The potential energy curve becomes lower for increasing strain.}
  \label{fig_potential_strain}
\end{figure}

\begin{figure}
  \begin{center}
    \scalebox{1.0}[1.0]{\includegraphics[width=8cm]{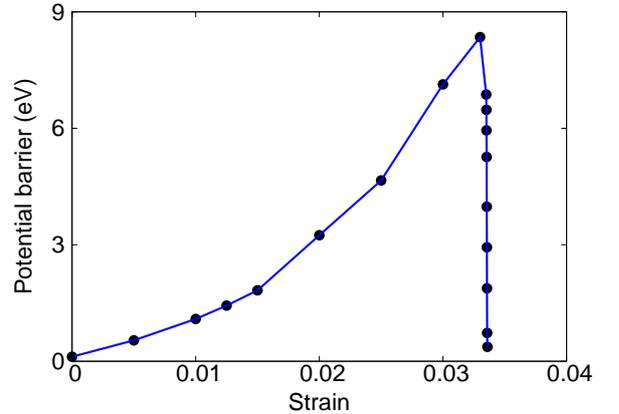}}
  \end{center}
  \caption{(Color online) The potential barrier $\Delta V$ for knotting at different compressive strains.}
  \label{fig_barrier}
\end{figure}

The potential energy curve of the knotting at different strains $\epsilon$ is displayed in Fig.~\ref{fig_potential_strain}. Fig.~\ref{fig_potential_strain}~(a) shows that the potential energy curve becomes higher for larger strain, when the applied compression is smaller than 0.033. In particular, the potential energy barrier $\Delta V$ in Fig.~\ref{fig_barrier} increases with increasing compression, so that it becomes more difficult to unknot the knot by applying strain. Fig.~\ref{fig_potential_strain}~(b) shows a quite different situation when the applied compressive strain is larger than 0.033, in which the potential energy curve decreases for increasing compression. In particular, Fig.~\ref{fig_barrier} shows that the potential energy barrier $\Delta V$ drops rapidly, and becomes almost zero at $\epsilon=0.03358$, so the structure can be deformed easily from the configuration with $\eta=-1$ to the configuration with $\eta=1$.  According to this $\eta$-potential argument, the knotting will be unknotted at strain $\epsilon=0.03358$, which is exactly the same as the critical unknotting strain $\epsilon_u$ determined by the stress-strain curve from MD simulations in Fig.~\ref{fig_width80}.

\subsection{Parametric Effects on Knotting}

We now perform a parametric analysis of the knotting effect, specifically taking into account the effects of graphene width, temperature, and orientation.

\begin{figure}
  \begin{center}
    \scalebox{1.2}[1.2]{\includegraphics[width=8cm]{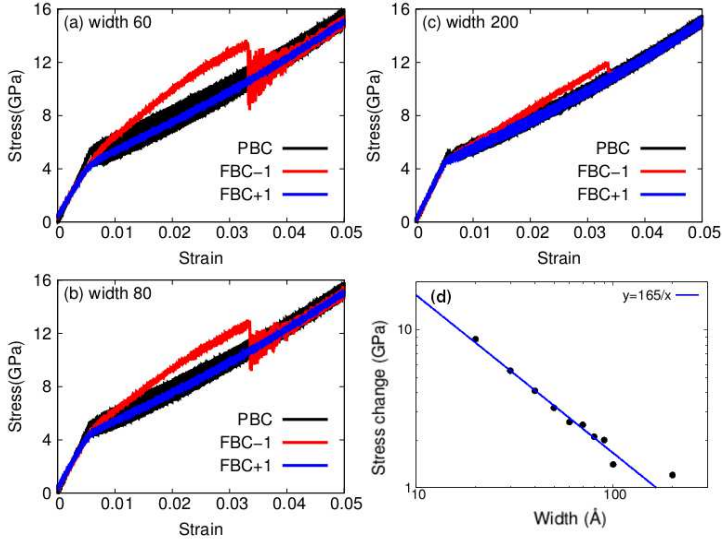}}
  \end{center}
  \caption{(Color online) Stress-strain for graphene of width (a) 60~{\AA}, (b) 80~{\AA}, and (c) 200~{\AA}. (d) The difference between the stress of the FBC-1 case and PBC case at the critical unknotting strain $\epsilon_u$.}
  \label{fig_all_width}
\end{figure}

Fig.~\ref{fig_all_width} shows the width dependence of the knotting effect on the buckling of graphene with length $L=30$~{\AA}.  Fig.~\ref{fig_all_width}~(d) shows that the difference ($\Delta \sigma$) between the maximum achievable stress after buckling for the FBC-1 case and the other two cases becomes smaller as the graphene width increases, and that the knotting effect is negligible in graphene with width 200~{\AA}. We can assume that graphene is divided into three regions: the two warped edge regions of width $l_{\rm eff}$ and one central region of width $W-2l_{\rm eff}$, with $l_{\rm eff}$ as the effective thickness for each edge region and $W$ as the total width. The stress difference $\Delta \sigma$ can be described by the formula, $\Delta \sigma=2(l_{\rm eff}/W) \Delta \sigma_E$, with $\Delta\sigma_E$ as the stress difference at the same strain between the edge region and the central region. From Fig.~\ref{fig_all_width}~(d), we have the fitted coefficient $2 l_{\rm eff}\Delta\sigma_E=165.2$. Using $l_c=7.3$~{\AA} as the effective thickness, i.e., $l_{\rm eff}=l_c=7.3$~{\AA}, it can be determined that $\Delta\sigma_E=11.3$~{GPa}. This value is slightly larger but close to the stress difference (8.7~{GPa}) for graphene of 20~{\AA} in width, which is dominated by the two edge regions. The two warped edge regions cause the buckling to be gradual for small widths, in contrast to the abrupt buckling of the central region for wider graphene.

\begin{figure}
  \begin{center}
    \scalebox{0.8}[0.8]{\includegraphics[width=8cm]{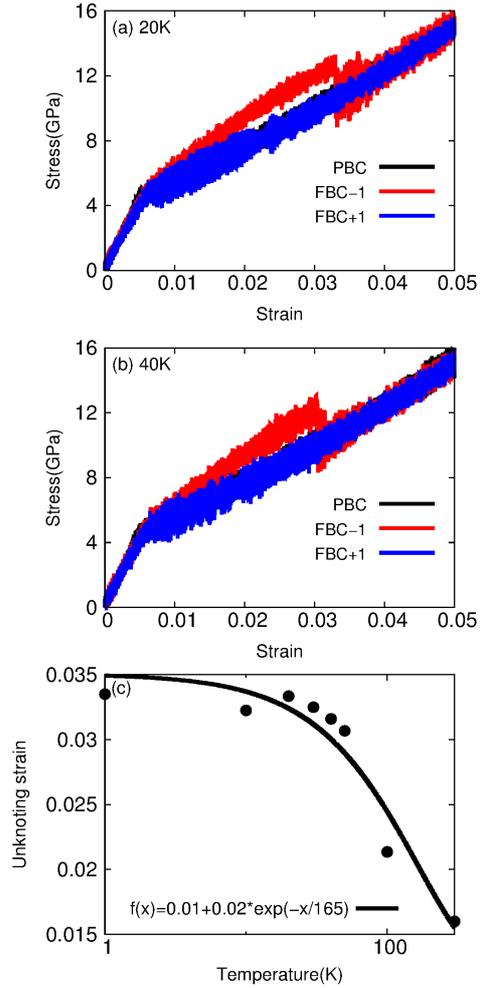}}
  \end{center}
  \caption{(Color online) Temperature effect on the knotting phenomenon for graphene of dimension $30\times 80$~{\AA}. The stress-strain relation for graphene at temperature (a) 20~K and (b) 40~K. (c) The temperature dependence for the unknotting strain, at which the knotting for graphene with FBC-1 configuration is unknotted.}
  \label{fig_temperature}
\end{figure}

Fig.~\ref{fig_temperature} shows the temperature dependence of the knotting effect on the buckling of graphene with dimension $30\times 80$~{\AA}.  In Fig.~\ref{fig_temperature}~(a), the knotting structure in graphene with FBC-1 configuration is unknotted at the critical unknotting strain $\epsilon_u=0.032$ at 20~K. The critical unknotting strain decreases to $\epsilon_u=0.0305$ at 40~K as shown in Fig.~\ref{fig_temperature}~(b), which indicates that the knotting structure is easier to be unknotted at higher temperature. It is because, at higher temperature, the thermal vibration energy is larger, so it is easier to overcome the potential energy barrier (in Fig.~\ref{fig_barrier}) of the knotting. Fig.~\ref{fig_temperature}~(c) shows the relation between temperature and the unknotting strain, which discloses an exponential decay of the unknotting strain with the increase of temperature.

\begin{figure}
  \begin{center}
    \scalebox{1.0}[1.0]{\includegraphics[width=8cm]{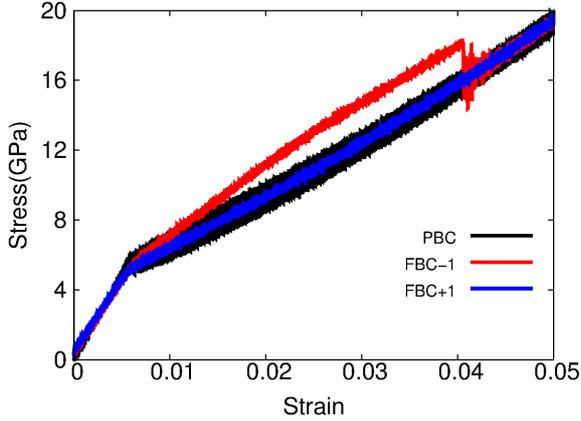}}
  \end{center}
  \caption{(Color online) Stress-strain for the compression of graphene along the zigzag orientation at 1.0~K. The dimension of the system is $30\times 80$~{\AA}.}
  \label{fig_zigzag}
\end{figure}

Finally, we discuss orientation effects on the knotting.  In the above, we have discussed the knotting effect on the buckling of graphene which is compressed along the armchair orientation. Fig.~\ref{fig_zigzag} shows that the knotting phenomenon can also be found in the buckling of graphene that is compressed along the zigzag orientation. This figure has similar features as that for the armchair graphene shown in Fig.~\ref{fig_width80}. The buckling process of graphene with FBC+1 configuration is similar as the buckling of graphene with PBC configuration. For graphene with FBC-1 configuration, the stress is obviously higher than the other two cases due to the knotting phenomenon.

\subsection{Randomness for Knotting Phenomenon}

\subsubsection{Width Dependence for Randomness}

\begin{figure}
  \begin{center}
    \scalebox{1.0}[1.0]{\includegraphics[width=8cm]{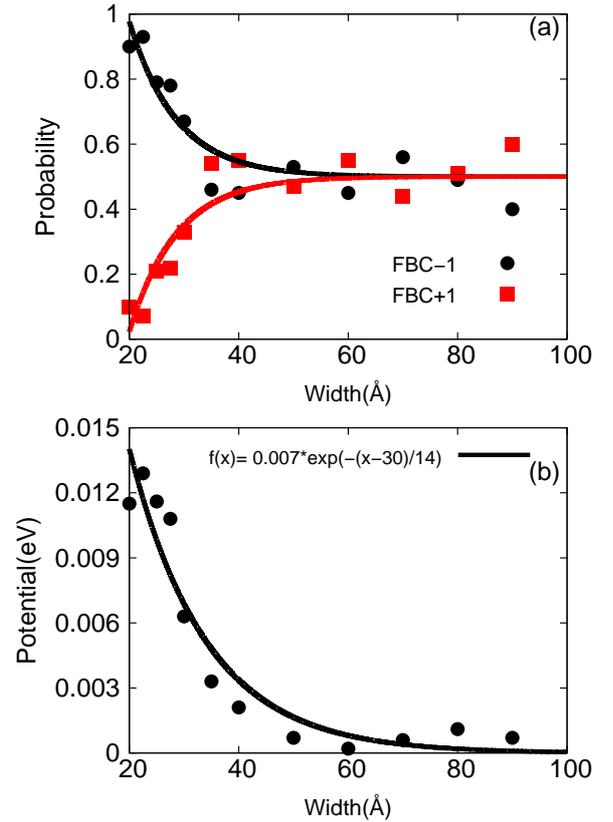}}
  \end{center}
  \caption{(Color online) Probability for FBC-1 and FBC+1 cases. (a) Width dependence for the probability of FBC-1 and FBC+1 cases in graphene with length $L=30$~{\AA} at 1.0~K. (b) The width dependence for the potential difference, $\Delta V=V_{\rm FBC-1}-V_{\rm FBC+1}$, between graphene with FBC-1 and FBC+1 configurations.}
  \label{fig_random_width}
\end{figure}

We have previously shown that for a free edge pair, each edge can be warped in the $\pm$z direction, resulting in the FBC-1 or FBC+1 configuration shown in Fig.~\ref{fig_sin_e_structure}. The warping direction of each isolated free edge can be either in the +z or -z direction with the same probability, because these two types of warped edges have the same potential energy.   On the one hand, if there is no coupling between the two free edges, a pair of free edges with FBC-1 configuration or FBC+1 configuration have the same potential energy, so the probabilities for the FBC-1 and the FBC+1 configurations are the same. On the other hand, if there is coupling between the two free edges, it is possible that graphene with FBC-1 configuration will have a different potential from the FBC+1 configuration, so the probability for FBC-1 and FBC+1 configurations will be different.

Indeed, Fig.~\ref{fig_random_width}~(a) shows that the probabilities for FBC-1 and FBC+1 configurations are width dependent at 1.0~K in graphene with FBC in the y-direction; i.e., with a pair of free edges in the y-direction. In this set of calculations, we perform thermalization for graphene with FBC in the y-direction within the NPT ensemble for 200~ps. The initial graphene structure is accompanied by a pair of free edges, but both edges are not warped at the initial stage. After thermalization, we find that both free edges are warped and the pair of free edges are either in the FBC-1 configuration or the FBC+1 configuration. We performed 100 simulations for the same graphene at each width, but with different initial random velocity distribution. After thermalization, we counted the number of the structure with FBC-1 configuration and the FBC+1 configuration, and the corresponding probabilities were calculated. We find that for narrow graphene, the probability for structure with FBC-1 configuration is obviously larger than the structure with FBC+1 configuration. This difference decreases with increasing width, and vanishes for width above 50~{\AA}.

The above probability results can be analyzed in terms of the potential energy difference between the structure with FBC-1 and FBC+1 configurations. Fig.~\ref{fig_random_width}~(b) shows the potential energy difference $\Delta V=V_{\rm FBC-1}-V_{\rm FBC+1}$ for graphene of different width. It shows that the potential for the FBC-1 configuration is lower than FBC+1 configuration especially for narrow graphene, which is the reason for the larger probability of graphene with FBC-1 than FBC+1 configuration in narrow graphene. For wide graphene, the potential difference becomes very small, so the probabilities for FBC-1 and FBC+1 configurations are almost the same. For wide graphene, two warped free edges are far from each other, so they can be regarded as isolated warped edges. As we know from Fig.~\ref{fig_potential_strain_0.0}, the potential energy is independent of the warping direction (upward or downward) in an isolated free edge, so the potential energy difference between FBC-1 and FBC+1 is almost zero for wide graphene, leading to the same probability of FBC-1 and FBC+1 configurations in wide graphene.

\begin{figure}
  \begin{center}
    \scalebox{1.0}[1.0]{\includegraphics[width=8cm]{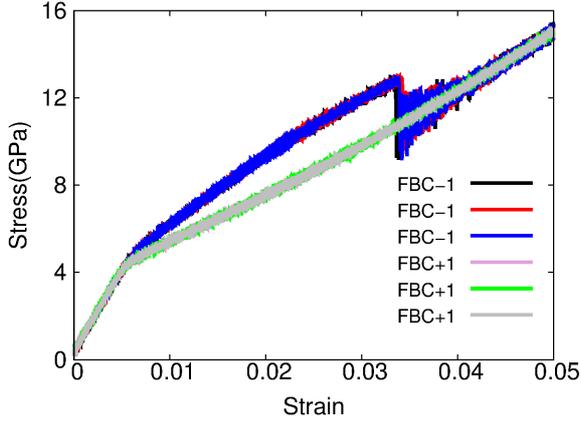}}
  \end{center}
  \caption{(Color online) Stress-strain for the compression of graphene at 1.0~K. FBC is applied in the y-direction. The dimension is $30\times 80$~{\AA}. The stress strain relation for graphene with FBC-1 configuration fall into the same curve, while the stress strain relation for graphene with FBC+1 configuration fall into another curve.}
  \label{fig_random_width80}
\end{figure}

The importance of the randomness is that most atomistic simulation studies start with a flat ideal initial graphene sheet with FBC, which will be thermalized to a stable structure at finite temperature. The resulting stable structure can be either FBC-1 or FBC+1 configuration with certain probability, which is width dependent as illustrated in Fig.~\ref{fig_random_width}~(a). Furthermore, Fig.~\ref{fig_random_width80} shows that the stress strain relations for all graphene with the FBC-1 configuration fall into one curve; while the stress strain relations for all graphene with the FBC+1 configuration fall into another curve. There is obvious difference between these two groups of stress-strain curves, which indicates that numerical simulations should obtain two different stress-strain relations for the same graphene, provided the free edges are not pre-warped in the initial structure.

\subsubsection{Temperature Dependence for Randomness}

\begin{figure}
  \begin{center}
    \scalebox{1.0}[1.0]{\includegraphics[width=8cm]{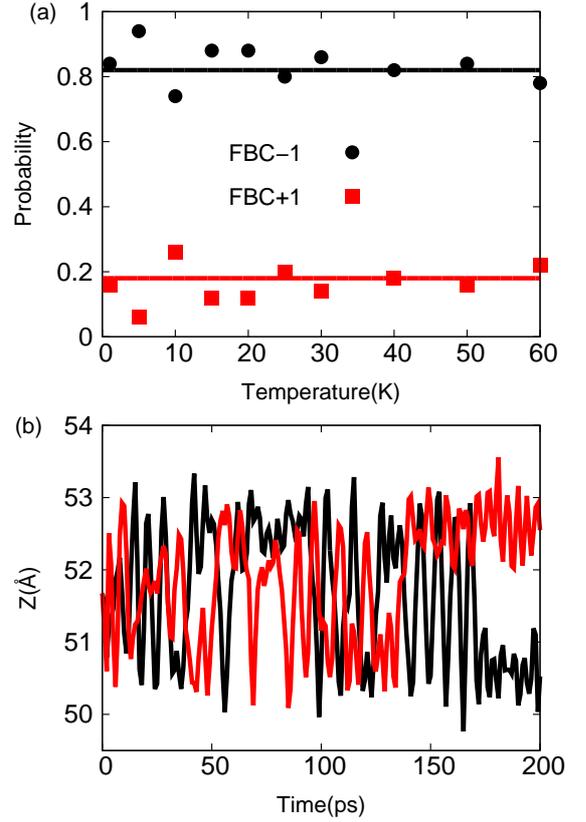}}
  \end{center}
  \caption{(Color online) Temperature dependence for the probability of FBC-1 and FBC+1 cases in graphene of width $W=20$~{\AA}. (a) The probability for FBC-1 configuration is always larger than FBC+1 configuration. (b) The z-position for the two warped edges, which shows the correlated flipping exhibited by the two edges, i.e. when one edge flips its warping direction, the other edge will flip its warping direction simultaneously.}
  \label{fig_random_temperature}
\end{figure}

We showed in Fig.~\ref{fig_random_width}~(a) that graphene with width $W=20$~{\AA} has a larger probability in the FBC-1 configuration than the FBC+1 configuration. We also showed in Fig.~\ref{fig_unstable_tem} that an isolated warped free edge has larger probability to flip its warping direction at higher temperatures.  Hence, it is natural to anticipate that the structure with FBC-1 configuration may be driven into the FBC+1 configuration by thermal vibrations at higher temperatures. In other words, it is expected that, for graphene with $W=20$~{\AA}, the probability of FBC-1 case will be reduced and becomes closer to the probability of FBC+1 case, if the temperature is increased. In this set of simulations, we initialized the velocity of the system with 50 different random velocity distributions for each temperature. The system was thermalized to its thermally stable structure within the NPT ensemble for 200~ps. After thermalization, both free edges in graphene are warped, and they are either in the FBC-1 configuration or in the FBC+1 configuration. The numbers for FBC-1 and FBC+1 cases were collected and their probabilities were calculated accordingly.

In Fig.~\ref{fig_random_temperature}~(a), the probability for graphene with FBC-1 configuration is always larger than graphene with FBC+1 configuration in the whole temperature range. For low temperatures, it is reasonable that the probability for graphene with FBC-1 configuration is larger than FBC+1 configuration, because we know that the potential for FBC-1 is lower than FBC+1 for graphene of width $W=20$~{\AA} in Fig.~\ref{fig_random_width}~(b). However, it is surprising that the probability for graphene with FBC-1 configuration is still larger than graphene with FBC+1 configuration at higher temperatures. This surprising result is attributed to the correlated flipping exhibited by the two free edges in Fig.~\ref{fig_random_temperature}~(b), which displays the z-position for the two warped free edges.  As can be seen, when the warping direction of one free edge is flipped, the warping direction of the other free edge also flips simultaneously.  This correlated flipping mechanism maintains the FBC-1 configuration, which ensures the larger probability for the FBC-1 configuration even at higher temperatures.

\section{Conclusion}

In conclusion, we have demonstrated a novel knotting phenomena induced by the interaction between free edges during the compression of graphene.  The knotting phenomenon has substantial effects on the mechanical properties of buckled graphene, in particular significantly elevating the stress that can be sustained during the buckling process, which results in a higher mechanical stiffness than graphene without knotting, and in enabling graphene to exhibit a stable post-buckling regime where the amount of strain that can be sustained is significantly larger than the pre-buckling elastic strain.  The knotting process was shown to be most probable for narrow graphene ribbons at lower temperatures.  Overall, we have shown that edge effects, which have previously been shown to cause undesired instabilities on the mechanical response of graphene, can be utilized to give surprising enhancements in mechanical performance.

\section{Acknowledgements} The work is supported by the Recruitment Program of Global Youth Experts of China, the National Natural Science Foundation of China (NSFC) under Grant Nos. 11504225, 11472163, 11425209, and the start-up funding from Shanghai University. HSP acknowledges the support of the Mechanical Engineering department at Boston University.


\begin{thebibliography}{38}
\providecommand{\natexlab}[1]{#1}
\providecommand{\url}[1]{\texttt{#1}}
\providecommand{\urlprefix}{URL }
\expandafter\ifx\csname urlstyle\endcsname\relax
  \providecommand{\doi}[1]{doi:\discretionary{}{}{}#1}\else
  \providecommand{\doi}{doi:\discretionary{}{}{}\begingroup
  \urlstyle{rm}\Url}\fi
\providecommand{\eprint}[2][]{\url{#2}}
\providecommand{\BIBand}{and}
\providecommand{\bibinfo}[2]{#2}
\ifx\xfnm\undefined \def\xfnm[#1]{\unskip,\space#1}\fi
\bibitem[{Lee et~al.(2008)Lee, Wei, Kysar and Hone}]{LeeC2008sci}
\bibinfo{author}{Lee\xfnm[ C.]}, \bibinfo{author}{Wei\xfnm[ X.]},
  \bibinfo{author}{Kysar\xfnm[ J.W.]}, \bibinfo{author}{Hone\xfnm[ J.]}.
\newblock \bibinfo{title}{Measurement of the elastic properties and intrinsic
  strength of monolayer graphene}.
\newblock \bibinfo{journal}{Science}
  \bibinfo{year}{2008};\bibinfo{volume}{321}:\bibinfo{pages}{385}.
\bibitem[{Ou-Yang. et~al.(1997)Ou-Yang., bin Su and Wang}]{OuyangZC1997}
\bibinfo{author}{Ou-Yang.\xfnm[ Z.C.]}, \bibinfo{author}{bin Su\xfnm[ Z.]},
  \bibinfo{author}{Wang\xfnm[ C.L.]}.
\newblock \bibinfo{title}{Coil formation in multishell carbon nanotubes:
  Competition between curvature elasticity and interlayer adhesion}.
\newblock \bibinfo{journal}{Physical Review Letters}
  \bibinfo{year}{1997};\bibinfo{volume}{78}(\bibinfo{number}{21}):\bibinfo{pages}{4055}.
\bibitem[{Tu and Ou-Yang(2002)}]{TuZC2002}
\bibinfo{author}{Tu\xfnm[ Z.C.]}, \bibinfo{author}{Ou-Yang\xfnm[ Z.C.]}.
\newblock \bibinfo{title}{Single-walled and multiwalled carbon nanotubes viewed
  as elastic tubes with the effective young’s moduli dependent on layer
  number}.
\newblock \bibinfo{journal}{Physical Review B}
  \bibinfo{year}{2002};\bibinfo{volume}{65}:\bibinfo{pages}{233407}.
\bibitem[{Arroyo and Belytschko(2004)}]{ArroyoM2004}
\bibinfo{author}{Arroyo\xfnm[ M.]}, \bibinfo{author}{Belytschko\xfnm[ T.]}.
\newblock \bibinfo{title}{Finite crystal elasticity of carbon nanotubes based
  on the exponential cauchy-born rule}.
\newblock \bibinfo{journal}{Physical Review B}
  \bibinfo{year}{2004};\bibinfo{volume}{69}:\bibinfo{pages}{115415}.
\bibitem[{Lu et~al.(2009)Lu, Arroyo and Huang}]{LuQ2009}
\bibinfo{author}{Lu\xfnm[ Q.]}, \bibinfo{author}{Arroyo\xfnm[ M.]},
  \bibinfo{author}{Huang\xfnm[ R.]}.
\newblock \bibinfo{title}{Elastic bending modulus of monolayer graphene}.
\newblock \bibinfo{journal}{Journal of Physics D: Applied Physics}
  \bibinfo{year}{2009};\bibinfo{volume}{42}:\bibinfo{pages}{102002}.
\bibitem[{Timoshenko and Woinowsky-Krieger(1987)}]{TimoshenkoS1987}
\bibinfo{author}{Timoshenko\xfnm[ S.]},
  \bibinfo{author}{Woinowsky-Krieger\xfnm[ S.]}.
\newblock \bibinfo{title}{Theory of Plates and Shells, 2nd ed}.
\newblock \bibinfo{publisher}{McGraw-Hill, New York}; \bibinfo{year}{1987}.
\bibitem[{Bao et~al.(2009)Bao, Miao, Chen, Zhang, Jang, Dames
  et~al.}]{BaoW2009nn}
\bibinfo{author}{Bao\xfnm[ W.]}, \bibinfo{author}{Miao\xfnm[ F.]},
  \bibinfo{author}{Chen\xfnm[ Z.]}, \bibinfo{author}{Zhang\xfnm[ H.]},
  \bibinfo{author}{Jang\xfnm[ W.]}, \bibinfo{author}{Dames\xfnm[ C.]}, et~al.
\newblock \bibinfo{title}{Controlled ripple texturing of suspended graphene and
  ultrathin graphite membranes}.
\newblock \bibinfo{journal}{Nature Nanotechnology}
  \bibinfo{year}{2009};\bibinfo{volume}{4}:\bibinfo{pages}{562--566}.
\bibitem[{Cong and Yu(2014)}]{CongC2014nc}
\bibinfo{author}{Cong\xfnm[ C.]}, \bibinfo{author}{Yu\xfnm[ T.]}.
\newblock \bibinfo{title}{Enhanced ultra-low-frequency interlayer shear modes
  in folded graphene layers}.
\newblock \bibinfo{journal}{Nature Communications}
  \bibinfo{year}{2014};\bibinfo{volume}{5}:\bibinfo{pages}{4709}.
\bibitem[{Lu and Huang(2009)}]{LuQ2009ijam}
\bibinfo{author}{Lu\xfnm[ Q.]}, \bibinfo{author}{Huang\xfnm[ R.]}.
\newblock \bibinfo{title}{Nonlinear mechanics of single-atomic-layer graphene
  sheets}.
\newblock \bibinfo{journal}{International Journal of Applied Mechanics}
  \bibinfo{year}{2009};\bibinfo{volume}{1}(\bibinfo{number}{3}):\bibinfo{pages}{443--467}.
\bibitem[{Patrick(2010)}]{PatrickWJ2010jctn}
\bibinfo{author}{Patrick\xfnm[ W.J.]}.
\newblock \bibinfo{title}{Buckling of graphene layers supported by rigid
  substrates}.
\newblock \bibinfo{journal}{Journal of Computational and Theoretical
  Nanoscience}
  \bibinfo{year}{2010};\bibinfo{volume}{7}(\bibinfo{number}{11}):\bibinfo{pages}{2338--2348}.
\bibitem[{Sakhaee-Pour(2009)}]{SakhaeePA2009cms}
\bibinfo{author}{Sakhaee-Pour\xfnm[ A.]}.
\newblock \bibinfo{title}{Elastic buckling of single-layered graphene sheet}.
\newblock \bibinfo{journal}{Computational Materials Science}
  \bibinfo{year}{2009};\bibinfo{volume}{45}(\bibinfo{number}{2}):\bibinfo{pages}{266--270}.
\bibitem[{Pradhan and Murmu(2009)}]{PradhanSC2009cms}
\bibinfo{author}{Pradhan\xfnm[ S.C.]}, \bibinfo{author}{Murmu\xfnm[ T.]}.
\newblock \bibinfo{title}{Small scale effect on the buckling of single-layered
  graphene sheets under biaxial compression via nonlocal continuum mechanics}.
\newblock \bibinfo{journal}{Computational Materials Science}
  \bibinfo{year}{2009};\bibinfo{volume}{47}(\bibinfo{number}{1}):\bibinfo{pages}{268--274}.
\bibitem[{Pradhan(2009)}]{PradhanSC2009plsa}
\bibinfo{author}{Pradhan\xfnm[ S.C.]}.
\newblock \bibinfo{title}{Buckling of single layer graphene sheet based on
  nonlocal elasticity and higher order shear deformation theory}.
\newblock \bibinfo{journal}{Physics Letters, Section A: General, Atomic and
  Solid State Physics}
  \bibinfo{year}{2009};\bibinfo{volume}{373}(\bibinfo{number}{45}):\bibinfo{pages}{4182--4188}.
\bibitem[{Frank et~al.(2010)Frank, Tsoukleri, Parthenios, Papagelis, Riaz,
  Jalil et~al.}]{FrankO2010acsnn}
\bibinfo{author}{Frank\xfnm[ O.]}, \bibinfo{author}{Tsoukleri\xfnm[ G.]},
  \bibinfo{author}{Parthenios\xfnm[ J.]}, \bibinfo{author}{Papagelis\xfnm[
  K.]}, \bibinfo{author}{Riaz\xfnm[ I.]}, \bibinfo{author}{Jalil\xfnm[ R.]},
  et~al.
\newblock \bibinfo{title}{Compression behavior of single-layer graphenes}.
\newblock \bibinfo{journal}{ACS Nano}
  \bibinfo{year}{2010};\bibinfo{volume}{4}(\bibinfo{number}{6}):\bibinfo{pages}{3131--3138}.
\bibitem[{Farajpour et~al.(2011)Farajpour, Mohammadi, Shahidi and
  Mahzoon}]{FarajpourA2011pe}
\bibinfo{author}{Farajpour\xfnm[ A.]}, \bibinfo{author}{Mohammadi\xfnm[ M.]},
  \bibinfo{author}{Shahidi\xfnm[ A.R.]}, \bibinfo{author}{Mahzoon\xfnm[ M.]}.
\newblock \bibinfo{title}{Axisymmetric buckling of the circular graphene sheets
  with the nonlocal continuum plate model}.
\newblock \bibinfo{journal}{Physica E: Low-dimensional Systems and
  Nanostructures}
  \bibinfo{year}{2011};\bibinfo{volume}{43}(\bibinfo{number}{10}):\bibinfo{pages}{1820--1825}.
\bibitem[{Tozzini and Pellegrini(2011)}]{TozziniV2011jpcc}
\bibinfo{author}{Tozzini\xfnm[ V.]}, \bibinfo{author}{Pellegrini\xfnm[ V.]}.
\newblock \bibinfo{title}{Reversible hydrogen storage by controlled buckling of
  graphene layers}.
\newblock \bibinfo{journal}{Journal of Physical Chemistry C}
  \bibinfo{year}{2011};\bibinfo{volume}{115}(\bibinfo{number}{51}):\bibinfo{pages}{25523--25528}.
\bibitem[{Rouhi and Ansari(2012)}]{RouhiS2012pe}
\bibinfo{author}{Rouhi\xfnm[ S.]}, \bibinfo{author}{Ansari\xfnm[ R.]}.
\newblock \bibinfo{title}{Atomistic finite element model for axial buckling and
  vibration analysis of single-layered graphene sheets}.
\newblock \bibinfo{journal}{Physica E: Low-dimensional Systems and
  Nanostructures}
  \bibinfo{year}{2012};\bibinfo{volume}{44}(\bibinfo{number}{4}):\bibinfo{pages}{764--772}.
\bibitem[{Giannopoulos(2012)}]{GiannopoulosGI2012cms}
\bibinfo{author}{Giannopoulos\xfnm[ G.I.]}.
\newblock \bibinfo{title}{Elastic buckling and flexural rigidity of graphene
  nanoribbons by using a unique translational spring element per interatomic
  interaction}.
\newblock \bibinfo{journal}{Computational Materials Science}
  \bibinfo{year}{2012};\bibinfo{volume}{53}(\bibinfo{number}{1}):\bibinfo{pages}{388--395}.
\bibitem[{Neek-Amal and Peeters(2012)}]{Neek-AmalM2012apl}
\bibinfo{author}{Neek-Amal\xfnm[ M.]}, \bibinfo{author}{Peeters\xfnm[ F.M.]}.
\newblock \bibinfo{title}{Effect of grain boundary on the buckling of graphene
  nanoribbons}.
\newblock \bibinfo{journal}{Applied Physics Letters}
  \bibinfo{year}{2012};\bibinfo{volume}{100}(\bibinfo{number}{10}):\bibinfo{pages}{101905}.
\bibitem[{Shen et~al.(2013)Shen, Xu and Zhang}]{ShenH2013apl}
\bibinfo{author}{Shen\xfnm[ H..]}, \bibinfo{author}{Xu\xfnm[ Y..]},
  \bibinfo{author}{Zhang\xfnm[ C..]}.
\newblock \bibinfo{title}{Graphene: Why buckling occurs?}
\newblock \bibinfo{journal}{Applied Physics Letters}
  \bibinfo{year}{2013};\bibinfo{volume}{102}(\bibinfo{number}{13}):\bibinfo{pages}{131905}.
\bibitem[{Jiang(2014)}]{JiangJW2014mos2buckling}
\bibinfo{author}{Jiang\xfnm[ J.W.]}.
\newblock \bibinfo{title}{The buckling of single-layer mos$_{2}$ under uniaxial
  compression}.
\newblock \bibinfo{journal}{Nanotechnology}
  \bibinfo{year}{2014};\bibinfo{volume}{25}:\bibinfo{pages}{355402}.
\bibitem[{Jiang(2015)}]{JiangJW2015reviewgramos2}
\bibinfo{author}{Jiang\xfnm[ J.W.]}.
\newblock \bibinfo{title}{Graphene versus mos2: A short review}.
\newblock \bibinfo{journal}{Front Phys}
  \bibinfo{year}{2015};\bibinfo{volume}{10}:\bibinfo{pages}{106801}.
\bibitem[{Brenner et~al.(2002)Brenner, Shenderova, Harrison, Stuart, Ni and
  Sinnott}]{brennerJPCM2002}
\bibinfo{author}{Brenner\xfnm[ D.W.]}, \bibinfo{author}{Shenderova\xfnm[
  O.A.]}, \bibinfo{author}{Harrison\xfnm[ J.A.]}, \bibinfo{author}{Stuart\xfnm[
  S.J.]}, \bibinfo{author}{Ni\xfnm[ B.]}, \bibinfo{author}{Sinnott\xfnm[
  S.B.]}.
\newblock \bibinfo{title}{A second-generation reactive empirical bond order
  ({REBO}) potential energy expression for hydrocarbons}.
\newblock \bibinfo{journal}{Journal of Physics: Condensed Matter}
  \bibinfo{year}{2002};\bibinfo{volume}{14}:\bibinfo{pages}{783--802}.
\bibitem[{Shenoy et~al.(2008)Shenoy, Reddy, Ramasubramaniam and
  Zhang}]{ShenoyVB}
\bibinfo{author}{Shenoy\xfnm[ V.B.]}, \bibinfo{author}{Reddy\xfnm[ C.D.]},
  \bibinfo{author}{Ramasubramaniam\xfnm[ A.]}, \bibinfo{author}{Zhang\xfnm[
  Y.W.]}.
\newblock \bibinfo{title}{Edge-stress-induced warping of graphene sheets and
  nanoribbons}.
\newblock \bibinfo{journal}{Physical Review Letters}
  \bibinfo{year}{2008};\bibinfo{volume}{101}(\bibinfo{number}{24}):\bibinfo{pages}{245501}.
\bibitem[{Gass et~al.(2008)Gass, Bangert, Bleloch, Wang, Nair and
  Geim}]{MhairiHG2008nn}
\bibinfo{author}{Gass\xfnm[ M.H.]}, \bibinfo{author}{Bangert\xfnm[ U.]},
  \bibinfo{author}{Bleloch\xfnm[ A.L.]}, \bibinfo{author}{Wang\xfnm[ P.]},
  \bibinfo{author}{Nair\xfnm[ R.R.]}, \bibinfo{author}{Geim\xfnm[ A.K.]}.
\newblock \bibinfo{title}{Free-standing graphene at atomic resolution}.
\newblock \bibinfo{journal}{Nature Nanotechnology}
  \bibinfo{year}{2008};\bibinfo{volume}{3}:\bibinfo{pages}{676--681}.
\bibitem[{Jia et~al.(2009)Jia, Hofmann, Meunier, Sumpter, Campos-Delgado,
  Romo-Herrera et~al.}]{JiaX2009sci}
\bibinfo{author}{Jia\xfnm[ X.]}, \bibinfo{author}{Hofmann\xfnm[ M.]},
  \bibinfo{author}{Meunier\xfnm[ V.]}, \bibinfo{author}{Sumpter\xfnm[ B.G.]},
  \bibinfo{author}{Campos-Delgado\xfnm[ J.]},
  \bibinfo{author}{Romo-Herrera\xfnm[ J.M.]}, et~al.
\newblock \bibinfo{title}{Controlled formation of sharp zigzag and armchair
  edges in graphitic nanoribbons}.
\newblock \bibinfo{journal}{Science}
  \bibinfo{year}{2009};\bibinfo{volume}{323}:\bibinfo{pages}{1701}.
\bibitem[{Engelund et~al.(2010)Engelund, Furst, Jauho and
  Brandbyge}]{EngelundM2010prl}
\bibinfo{author}{Engelund\xfnm[ M.]}, \bibinfo{author}{Furst\xfnm[ J.A.]},
  \bibinfo{author}{Jauho\xfnm[ A.P.]}, \bibinfo{author}{Brandbyge\xfnm[ M.]}.
\newblock \bibinfo{title}{Localized edge vibrations and edge reconstruction by
  joule heating in graphene nanostructures}.
\newblock \bibinfo{journal}{Physical Review Letters}
  \bibinfo{year}{2010};\bibinfo{volume}{104}:\bibinfo{pages}{036807}.
\bibitem[{Kim and Park(2009)}]{KimSY2009nl}
\bibinfo{author}{Kim\xfnm[ S.Y.]}, \bibinfo{author}{Park\xfnm[ H.S.]}.
\newblock \bibinfo{title}{The importance of edge effects on the intrinsic loss
  mechanisms of graphene nanoresonators}.
\newblock \bibinfo{journal}{Nano Letters}
  \bibinfo{year}{2009};\bibinfo{volume}{9}(\bibinfo{number}{3}):\bibinfo{pages}{969--974}.
\bibitem[{Jiang and Wang(2012)}]{JiangJW2012jap}
\bibinfo{author}{Jiang\xfnm[ J.W.]}, \bibinfo{author}{Wang\xfnm[ J.S.]}.
\newblock \bibinfo{title}{Why edge effects are important on the intrinsic loss
  mechanisms of graphene nanoresonators}.
\newblock \bibinfo{journal}{Journal of Applied Physics}
  \bibinfo{year}{2012};\bibinfo{volume}{111}(\bibinfo{number}{5}):\bibinfo{pages}{054314}.
\bibitem[{Guo et~al.(2011)Guo, Chang, Guo and Gao}]{GuoZ2011prl}
\bibinfo{author}{Guo\xfnm[ Z.]}, \bibinfo{author}{Chang\xfnm[ T.]},
  \bibinfo{author}{Guo\xfnm[ X.]}, \bibinfo{author}{Gao\xfnm[ H.]}.
\newblock \bibinfo{title}{Thermal-induced edge barriers and forces in
  interlayer interaction of concentric carbon nanotubes}.
\newblock \bibinfo{journal}{Physical Review Letters}
  \bibinfo{year}{2011};\bibinfo{volume}{107}(\bibinfo{number}{10}):\bibinfo{pages}{105502}.
\bibitem[{Chang et~al.(2015)Chang, Zhang, Guo, Guo and Gao}]{ChangT2015prl}
\bibinfo{author}{Chang\xfnm[ T.]}, \bibinfo{author}{Zhang\xfnm[ H.]},
  \bibinfo{author}{Guo\xfnm[ Z.]}, \bibinfo{author}{Guo\xfnm[ X.]},
  \bibinfo{author}{Gao\xfnm[ H.]}.
\newblock \bibinfo{title}{Nanoscale directional motion towards regions of
  stiffness}.
\newblock \bibinfo{journal}{Physical Review Letters}
  \bibinfo{year}{2015};\bibinfo{volume}{114}(\bibinfo{number}{1}):\bibinfo{pages}{015504}.
\bibitem[{Castro~Neto et~al.(2009)Castro~Neto, Guinea, Peres, Novoselov and
  Geim}]{CastroNAH}
\bibinfo{author}{Castro~Neto\xfnm[ A.H.]}, \bibinfo{author}{Guinea\xfnm[ F.]},
  \bibinfo{author}{Peres\xfnm[ N.M.R.]}, \bibinfo{author}{Novoselov\xfnm[
  K.S.]}, \bibinfo{author}{Geim\xfnm[ A.K.]}.
\newblock \bibinfo{title}{The electronic properties of graphene}.
\newblock \bibinfo{journal}{Rev Mod Phys}
  \bibinfo{year}{2009};\bibinfo{volume}{81}(\bibinfo{number}{1}):\bibinfo{pages}{109--162}.
\bibitem[{Nose(1984)}]{Nose}
\bibinfo{author}{Nose\xfnm[ S.]}.
\newblock \bibinfo{title}{A unified formulation of the constant temperature
  molecular dynamics methods}.
\newblock \bibinfo{journal}{Journal of Chemical Physics}
  \bibinfo{year}{1984};\bibinfo{volume}{81}(\bibinfo{number}{1}):\bibinfo{pages}{511}.
\bibitem[{Hoover(1985)}]{Hoover}
\bibinfo{author}{Hoover\xfnm[ W.G.]}.
\newblock \bibinfo{title}{Canonical dynamics: Equilibrium phase-space
  distributions}.
\newblock \bibinfo{journal}{Physical Review A}
  \bibinfo{year}{1985};\bibinfo{volume}{31}(\bibinfo{number}{3}):\bibinfo{pages}{1695}.
\bibitem[{Plimpton(1995)}]{PlimptonSJ}
\bibinfo{author}{Plimpton\xfnm[ S.J.]}.
\newblock \bibinfo{title}{Fast parallel algorithms for short-range molecular
  dynamics}.
\newblock \bibinfo{journal}{Journal of Computational Physics}
  \bibinfo{year}{1995};\bibinfo{volume}{117}:\bibinfo{pages}{1--19}.
\bibitem[{Lammps(2012)}]{Lammps}
\bibinfo{author}{Lammps\xfnm[]}.
\newblock \bibinfo{journal}{http://wwwcssandiagov/$\sim$sjplimp/lammpshtml}
  \bibinfo{year}{2012};.
\bibitem[{Stukowski(2010)}]{ovito}
\bibinfo{author}{Stukowski\xfnm[ A.]}.
\newblock \bibinfo{title}{Visualization and analysis of atomistic simulation
  data with ovito - the open visualization tool}.
\newblock \bibinfo{journal}{Modelling and Simulation in Materials Science and
  Engineering}
  \bibinfo{year}{2010};\bibinfo{volume}{18}:\bibinfo{pages}{015012}.
\bibitem[{Coulais et~al.(2015)Coulais, Overvelde, Lubbers, Bertoldi and van
  Hecke}]{CoulaisC2015prl}
\bibinfo{author}{Coulais\xfnm[ C.]}, \bibinfo{author}{Overvelde\xfnm[ J.T.]},
  \bibinfo{author}{Lubbers\xfnm[ L.A.]}, \bibinfo{author}{Bertoldi\xfnm[ K.]},
  \bibinfo{author}{van Hecke\xfnm[ M.]}.
\newblock \bibinfo{title}{Discontinuous buckling of wide beams and metabeams}.
\newblock \bibinfo{journal}{Physical Review Letters}
  \bibinfo{year}{2015};\bibinfo{volume}{115}(\bibinfo{number}{4}):\bibinfo{pages}{044301}.

\end{thebibliography}

\end{document}